\documentclass[aps,prl,amssymb,showpacs,twocolumn]{revtex4}
\usepackage{graphicx}

\newcommand{\Case}[2]{{\textstyle \frac{#1}{#2}}}
\newcommand{\lP}{\ell_{\mathrm P}}

\begin{document}

\preprint{IMSc/2005/4/11}

\title{Absence of the Kasner singularity in the effective dynamics from 
loop quantum cosmology}

\author{Ghanashyam Date}
\email{shyam@imsc.res.in}
\affiliation{The Institute of Mathematical Sciences,
CIT Campus, Chennai-600 113, INDIA}


\begin{abstract} 

In classical general relativity, the generic approach to the initial
singularity is usually understood in terms of the BKL scenario. In this
scenario, along with the Bianchi IX model, the exact, singular, Kasner
solution of vacuum Bianchi I model also plays a pivotal role. Using an
effective classical Hamiltonian obtained from loop quantization of
vacuum Bianchi I model, exact solution is obtained which is non-singular
due to a discreteness parameter. The solution is parameterized in
exactly the same manner as the usual Kasner solution and reduces to the
Kasner solution as discreteness parameter is taken to zero. At the
effective Hamiltonian level, the avoidance of Kasner singularity uses a
mechanism distinct from the `inverse volume' modifications
characteristic of loop quantum cosmology.

\end{abstract}

\maketitle

While the celebrated singularity theorems of classical general
relativity imply that the backward evolution of an expanding universe
leads to a singular state, the {\em nature of the singularity} is
elucidated in terms of the Belinskii-Khalatnikov-Lifschitz (BKL)
scenario \cite{BKL,Rendall}. The scenario views the spatial slice close
to the (space-like) singularity as made up of approximately homogeneous
patches each of which evolves according to the vacuum Bianchi IX model.
As the singularity is approached, the patches fragment indefinitely,
asymptotically becoming infinitely small. The vacuum Bianchi IX
evolution in turn, can be viewed as a succession of Bianchi I evolutions
(Kasner epochs), interleaved by transitions among the Kasner epochs.
This interleaved evolution continues indefinitely in a chaotic manner.
Apart from the indefinite number of Kasner epochs and transitions among
these, the {\em monotonic decrease} of the volume and  the {\em
singular} nature of the Kasner solution are responsible for the infinite
fragmentation of the homogeneous patches.

Singularities `evolving' from non-singular physical situations are
indicative of breakdown of the extrapolation of the classical evolution
i.e. dynamics of Einstein equation and call far an
extension/modification of the classical theory/framework. One natural
avenue is to appeal to a corresponding quantum theory of gravity. Since
the classical picture of the singular behaviour involves highly
dynamical geometries with arbitrarily large curvatures, a quantum theory
which does {\em not} depend on any pre-selected background geometry is 
likely to be most suitable for obtaining the required extensions. 

Loop quantum gravity (LQG) is precisely such a background independent
approach \cite{LQGRev}. While the LQG for the general inhomogeneous
situations is quite complicated, its methods can be implemented and
tested in simpler contexts of spatially homogeneous geometries. Indeed,
loop quantization of the so-called diagonalized, Bianchi class A models
has been carried out \cite{HomCosmo,Spin} and shown to be non-singular
within the quantum framework. In the quantum framework, non-singularity
means non-breakdown of the fundamental dynamical equations (which are
partial difference equations) and boundedness of relevant operators such
as the inverse triad operator which enter the quantization of curvature
components, matter densities etc. 

It is of course more intuitive and convenient to obtain the {\em
modifications} implied by the quantum theory in the familiar geometrical
setting of classical  general relativity i.e. obtaining the
modifications to the Einstein {\em dynamics} keeping the kinematical
framework of Riemannian geometry in tact. This has been done
systematically for the isotropic models
\cite{SemiClass,EffHamiltonian,DiscretenessCorrections} in terms of an
effective Hamiltonian. The derivation of the effective Hamiltonian is
based on the observation that {\em if} the fundamental dynamical
equations admits a solution which is WKB approximable (i.e. the
amplitude and the phase are slowly varying in a suitable sense) in {\em
some} domain, {\em then} to the order $\hbar^0$ one obtains a
Hamilton-Jacobi equation from which an effective {\em classical}
Hamiltonian can be read-off, also valid within the same domain.  The
{\em largest possible} domain of validity of such an approximation is
constrained by the classical `turning points' dictated by the effective
Hamiltonian. The {\em form} of the effective Hamiltonian so obtained
does {\em not} depend on details of the presumed solution, but of course
the {\em actual} domain of validity of the WKB approximation and
consequently of the effective Hamiltonian is sensitive to the presumed
solution. Being $o(\hbar^0)$, the effective Hamiltonian is insensitive
to factor ordering issues.

There are various types of corrections that arise. The most dramatic one
is the correction implied by the non-trivial quantization of inverses of
various classical quantities such as scale factors, triad components,
volume etc \cite{InvScale,ICGCInv}. In the isotropic context, this has
lead to a variety of implications
\cite{Inflation,Bounce,DensityPert,PowerSuppression}.  For the
anisotropic context, this leads to suppression of chaotic behaviour of
the Bianchi IX model \cite{Chaos} with the further result that
asymptotically for vanishing volumes, a Bianchi IX solution approaches a
Kasner solution in a stable manner \cite{BIX}.

For the vacuum Bianchi I model, there is no potential term and no scope
for a modification of dynamics due to quantizations of inverses of
triad components. Although at the quantum level, there is no singularity
\cite{HomCosmo}, the effective classical Hamiltonian derived from a {\em
continuum approximation}, is identical to the Einstein Hamiltonian
leading to the same singular Kasner solution.  Recently however an
alternative method of deriving effective classical Hamiltonian has been
devised, in the context of isotropic models
\cite{DiscretenessCorrections}, which uses the WKB ansatz directly at
the difference equation level bypassing the step of first deriving the
Wheeler-De Witt differential equation to be followed by WKB
approximation. The same method can also be applied in the anisotropic
context which leads to an effective Hamiltonian {\em different} from the
Einstein Hamiltonian. Exact, non-singular solution of this Hamiltonian
for the vacuum Bianchi I model is the result presented here.

A Bianchi I space-time is specified in terms of the metric of the form,
\begin{equation} 
ds^2 = dt^2 - \sum_I a_I^2(t) (dx^I)^2 \ ,
\end{equation}
where $t$ is the synchronous time. The vacuum Einstein equations then
lead to the well known Kasner solution: $a_I(t) \sim t^{2 \alpha_I}$
where $\alpha_I$ are constants satisfying the conditions $\sum_I
\alpha_I^2 = 1 = \sum_I \alpha_I$. For subsequent comparison with
the new solution, it is convenient to describe the time evolution in
terms of a new time coordinate $\tau$ corresponding to the lapse ${\cal
N} := a_1 a_2 a_3$, defined by ${\cal N} d\tau = dt$.  The scale factors
then evolve as $a_I \sim e^{\alpha_I \tau}$.

Loop quantization of all diagonalized, Bianchi class A models has been
given in \cite{Spin}. Briefly, it may be summarized as follows. The
kinematical Hilbert space is spanned by orthonormalized vectors labeled
as $|\mu_1, \mu_2, \mu_3 \rangle, ~\mu_I \in \mathbb{R}$. These are
properly normalized eigenvectors of the triad operators $p^I$ with
eigenvalues $\Case{1}{2} \gamma \lP^2 \mu_I$, where $\gamma$ is the
Barbero-Immirzi parameter and $\lP^2 := 8 \pi G \hbar := \kappa \hbar$.
The volume operator is also diagonal in these labels with eigenvalues
$V(\vec{\mu})$ given by $(\Case{1}{2}\gamma\lP^2)^{3/2} \sqrt{|\mu_1
\mu_2 \mu_3|}$.  Here we have used the vector notation to denote the
triple $(\mu_1, \mu_2, \mu_3)$.  Imposing the Hamiltonian constraint
operator on general vectors of the form $|s\rangle = \sum_{\vec{\mu}}
s(\vec{\mu}) |\vec{\mu}\rangle$ leads to the fundamental difference
equation for the coefficients $s(\vec{\mu})$. Here the sum is over
countable subsets of $\mathbb{R}^3$. There are further gauge invariance
conditions \cite{HomCosmo} which do not concern us here. 

In the present context of vacuum Bianchi I model, the fundamental difference
equation takes the form \cite{Spin},
\begin{equation} 
\sum_{\vec{\epsilon}_{12}} A_{12}(\vec{\mu}; \vec{\epsilon}_{12})
s(\vec{\mu}; \vec{\epsilon}_{12}) + \text{cyclic} = 0\ , \text{where}
\end{equation} 
$\vec{\epsilon}_{12} = (\epsilon_1, \epsilon_2, \epsilon'_1,
\epsilon'_2)$ with each of the $\epsilon_*$ taking values $\pm1$;
$s(\vec{\mu}; \vec{\epsilon}_{12}) = s(\mu_1 - \mu_0 \epsilon_1 - \mu_0
\epsilon'_1, \mu_2 - \mu_0 \epsilon_2 - \mu_0 \epsilon'_2, \mu_2)$;
$\mu_0$ is an order 1 parameter and,
\begin{eqnarray}
A_{12}(\vec{\mu}; \vec{\epsilon}_{12}) & = &
V(\vec{\mu}; \vec{\epsilon}_{12}) d(\mu_3)( \epsilon_1 \epsilon_2 + \epsilon'_1
\epsilon'_2) \\
d(\mu) & :=  & \left\{\begin{array}{cl} 
\sqrt{|1 + \mu_0\mu^{-1}|} - \sqrt{|1 - \mu_0\mu^{-1}|}  &  
\mu \ne 0 \\
0 & \mu = 0 
\end{array} \right. \nonumber
\end{eqnarray}
In the summary above, we have used the non-separable kinematical Hilbert
space and also made the parameter $\mu_0$ explicit \cite{Bohr}.

To derive the effective Hamiltonian, one assumes that there exist
solution(s) of the partial difference equation which have a slowly
varying amplitude and phase at least in some region of large volume.
Explicitly, defining $\vec{p}(\mu_I) := \Case{1}{2} \gamma \lP^2
\vec{\mu}$, one introduces an interpolating function, $\psi(p)$, such
that $s(\vec{\mu}) = \psi(\vec{p}(\mu_I)) = C(\vec{p})$
exp$\{\Case{i}{\hbar} \Phi(\vec{p})\} $ and assumes that the amplitude
and phase are slowly varying functions of $\vec{p}$ in the sense that
higher order terms in the Taylor series about any $\vec{p}$ in the
relevant region, are smaller than the lower order terms when compared
over the {\em quantum geometry scale} $q := \Case{1}{2} \gamma \mu_0
\lP^2$.  Taylor expanding the interpolating wave function
$\psi(\vec{p};\vec{\epsilon}_{IJ}), \ (IJ) = (12, 23, 31)$, it is
straightforward to check that the leading terms (in powers of $\hbar$)
in the real part of the equation are $o(\hbar^0)$ while those in the
imaginary part are $o(\hbar)$. The $o(\hbar^0)$ terms involve only the
{\em first} order partial derivatives of the WKB phase. Identifying $K_1
:= \kappa \Phi_{1,0,0}(\vec{p})$ etc., one infers the Hamiltonian system
from the Hamilton-Jacobi equation as: $\{p^I, K_J\} = \kappa
\delta^I_J$, with the Hamiltonian ($\epsilon := \mu_0 \gamma$),
\begin{eqnarray}\label{VacBI}
\kappa {\cal N} H_{\text{eff}}(\vec{p}, \vec{K}) & = & - 2\left[ p^1 p^2
\frac{\text{sin} \epsilon K_1}{\epsilon} \frac{\text{sin} \epsilon
K_2}{\epsilon} + \text{cyclic} \right]
\end{eqnarray}
$\epsilon$ will be referred to as the discreteness parameter. 

The effective Hamiltonian is periodic in $\epsilon K_I$ and we may
restrict our attention to $- \pi < \epsilon K_I < \pi$. The imaginary
part of the equation however requires the domain of validity to be
restricted further in order to be self consistent with the assumption of
slow variation of the interpolating wave function. The restriction is:
$p_I \ge q $ and small neighbourhoods of $\pm \Case{\pi}{2}$ are to be
excluded. Thus along the $\epsilon K_I$ axes, the effective Hamiltonian
is a good approximation in the intervals: $(- \pi, - \Case{\pi}{2} -
\delta), (- \Case{\pi}{2} + \delta,  \Case{\pi}{2}  - \delta), (\Case{\pi}{2}
+ \delta, \pi)$ for some small positive $\delta$. The values $\epsilon
K_I = \pm \Case{\pi}{2}$ will turn out to be the `turning points' of the
trajectories, $\vec{p}(\tau)$, of the effective dynamics.

For $\epsilon K_I \sim 0$, one can use $\epsilon^{-1} \text{sin}\
\epsilon K_I \approx K_I$. Then the effective Hamiltonian goes over to
the Einstein Hamiltonian and the dependence on $\epsilon$ drops out.
Keeping only the $\hbar^0$ terms, we have effectively taken $|p^I|$
larger than the quantum geometry scale set by $q$. Thus it is clear that
Einstein dynamics is reproduced for large values of triads and small
values of their conjugates $K_I$. In this regime, the Hamilton's
equations for $p^I$, combined with the relation of the triad components
to scale factors, $|p^I| = a_J a_K$ (and cyclic permutations),
identifies the $K_I$'s as components of the extrinsic curvature of the
constant $t$ slices: $K_I = - \Case{1}{2} \Case{d a_I}{d t}$. The
effective Hamiltonian thus deviates from the Einstein Hamiltonian mainly
in the region of the phase space with not too small values of $\epsilon
K_I$. 

The small value of $\epsilon K_I$ can be achieved by taking the limit
$\epsilon \to 0$ ($K_I$ fixed) which then removes any restriction on
$K_I$'s. From a purely classical perspective, such a view may be welcome
especially since the parameters $\mu_0, \gamma$ disappear which are
absent in the Einstein dynamics. However, from a quantum perspective, the
discreteness parameter {\em cannot} be zero \cite{Bohr}.  For a non-zero
value of discreteness parameter, the effective Hamiltonian represents an
extension of Einstein dynamics for phase space regions beyond small
extrinsic curvatures and large triad components.  Since the non-zero
value of $\epsilon$ reflects a discrete structure in a specific
technical sense, the effective dynamics is to be viewed as a way of
extending the Einstein dynamics by incorporating the discrete feature of
quantum geometry. 

We use geometrized units ($\kappa = 1 = c$) so that all quantities have
dimensions of powers of length.  The triad variables $p^I$ and $\hbar$
have dimensions of (length)$^2$; the effective Hamiltonian, scale
factors, the synchronous time all have dimensions of length while the
$\tau$ has dimensions of (length)$^{-2}$. Using the naturally available
quantum geometry length scale of $\sqrt{q}$, {\em all dimensionfull
quantities below, are rendered dimensionless}. We will continue to use
the same symbols though.

The Hamilton's equations from the effective Hamiltonian are,
\begin{eqnarray}
\frac{d p^I}{d \tau} & = & -2 p^I \text{cos} \epsilon K_I \left(
\frac{p^J \text{sin} \epsilon K_J}{\epsilon} +
\frac{p^K \text{sin} \epsilon K_K}{\epsilon} \right) \ , \label{PEqn}\\
\frac{d K_I}{d \tau} & = & +2 \frac{\text{sin} \epsilon K_I}{\epsilon}
\left(
\frac{p^J \text{sin} \epsilon K_J}{\epsilon} +
\frac{p^K \text{sin} \epsilon K_K}{\epsilon} \right) \ , \label{KEqn}\\
0 & = & \left( \frac{p^1 \text{sin} \epsilon K_1}{\epsilon} \right)
\left( \frac{p^2 \text{sin} \epsilon K_2}{\epsilon} \right) ~+~
\text{cyclic} \ \label{ConstraintEqn}.
\end{eqnarray}

It follows immediately that $\Case{p^I \text{sin} \epsilon
K_I}{\epsilon} := -\Case{\alpha_I}{2}$ are constants of motion, with
$\left(\alpha_1 \alpha_2 + \text{cyclic} \right) = 0 = \sum_I \alpha_I^2
- \left(\sum_I \alpha_I \right)^2$ following from (\ref{ConstraintEqn}).

If all the $\alpha_I$ are zero, then all the $p^I, K_I$ are also
constants. Since lapse is non-zero, the $p^I$ are non-zero constants and
the solution represents the usual Minkowski space-time. The effective
dynamics thus retains the Minkowski space-time as a solution indicating
a good classical limit of the quantum dynamics. For a non-trivial
solution, then, $\sum_I \alpha_I^2 \ne 0$ must hold and as usual, by a
constant scaling of the ${\cal N}$ (or of $\tau$), one can arrange the
$\alpha_I$ to satisfy: $\sum_I \alpha_I = 1$. Thus the the {\em one
dimensional} parameter space of these solutions is exactly same as that
of the usual Kasner solutions.  (The special cases of the form $\alpha_1
= 1, \alpha_2 = \alpha_3 = 0$ are independent of $\epsilon$ in their
behaviour and are not considered here. Thus all $\alpha_I$ are assumed
to be non-zero.) It follows that exactly one of the $\alpha_I$'s must be
strictly negative and the remaining two strictly positive and 
$\alpha_J + \alpha_K = 1 - \alpha_I > 0$. 

Since $\alpha_I$ are non-zero constants, neither $p^I$ nor sin$\epsilon
K_I$ can vanish and thus cannot change sign. For definiteness, let us
restrict our attention to `positively oriented' triad and in particular
take $p^I > 0, \forall I$. Then, sgn(sin$\epsilon K_I$) = -
sgn($\alpha_I$), which fixes the two quadrants to which the `angles'
$\theta_I := \epsilon K_I$ must be confined along a solution i.e. either
$0 < \epsilon K_I < \pi$ or $- \pi < \epsilon K_I < 0$. Clearly, $p^I$
approach $\infty$ as $\epsilon K_I$ approach the end points of the
intervals and take the {\em minimum} value $\Case{\epsilon
|\alpha_I|}{2}$ for $\epsilon K_I = \pm \Case{\pi}{2}$. The traversal of
$\theta_I$ with $\tau$ is also fixed by (\ref{KEqn}): $\theta_I$ must
decrease if sin$\theta_I$ is positive and increase if sin$\theta_I$ is
negative i.e. we must have $\theta_I \to 0_{\pm}$ as $\tau \to \infty$ 
($\theta_I \to \pm \pi$ as $\tau \to - \infty$).

Eliminating $K_I$ in favour of $\alpha_I, p^I$, leaves us with an
equation for $p^I$, namely, 
\begin{equation}
\frac{d p^I}{d \tau} ~=~ \pm (1 - \alpha_I)
\sqrt{(p^I)^2 - \left(\frac{\epsilon \alpha_I}{2} \right)^2 } \ .
\end{equation}
The $\pm$ is determined by the quadrant to which the angle $\epsilon
K_I$ belongs. The solution is easily obtained as, 
\begin{equation}
p^I(\tau) = \epsilon \frac{|\alpha_I|}{2} \ 
\text{cosh}\left\{ (1 - \alpha_I)(\tau - (\tau_I)_0) \right\} \ .
\end{equation}
Notice that a triad component attains its smallest value,
$\Case{\epsilon |\alpha_I|}{2}$, at $\tau = (\tau_I)_0$ while for large
$|\tau|$ it behaves as $p^I \sim (\epsilon |\alpha_I|/4) \text{exp}( (1
- \alpha_I)|\tau| )$.

In terms of the scale factors, $a_I = \Case{\text{volume}}{p^I}$, the
solution is given by,
\begin{eqnarray}
a^2_I(\tau) & = & \epsilon \frac{1 - \alpha_I}{2} 
\left[\frac{}{}
\left(\text{cosh}\left\{ (1 - \alpha_J)(\tau - (\tau_J)_0)
\right\}\right) \right. \times \nonumber \\
& & \hspace{1.5cm}\left(\text{cosh}\left\{ (1 - \alpha_K)(\tau - (\tau_K)_0)
\right\}\right)  \times \nonumber \\
& & \hspace{1.4cm} \left.
\left({\text{cosh}\left\{ (1 - \alpha_I)(\tau - (\tau_I)_0)\right\} }
\right)^{-1}
\right]
\end{eqnarray}

For comparison, the triad and the scale factor for the Kasner solution are,
\begin{equation}
p^I(\tau) = p^I_0 e^{(1 - \alpha_I)\tau} ~~~,~~~
a^2_I(\tau) = (a_I)^2_0 e^{2 \alpha_I \tau} \ .
\end{equation}
One can recover the Kasner solution from the modified one by taking the
limit $(\tau_I)_0 \to - \infty, \epsilon \to 0$ such that $\epsilon
\Case{|\alpha_I|}{4} e^{-(1 - \alpha_I)(\tau_I)_0} = p^I_0$. 

For the Kasner solution, two scale factors vanish and third one
diverges such that the volume vanishes exponentially with $\tau$, as
$\tau \to - \infty$.  This translates into a {\em finite synchronous
time} $t$ in the past, making the Kasner solution singular.  

By contrast, for the modified solution, none of the triad variables can
become zero at any $\tau$ and the {\em volume never vanishes}.
Furthermore, due to the hyperbolic cosine function, for {\em both}
asymptotic times, the modified solution approaches the large volume
behaviour of the Kasner solution.  In particular, the scale factor
behaves as $a^2_I \to (\epsilon (1 - \alpha_I)/4) \text{exp} (2 \alpha_I
|\tau|)$ as $|\tau| \to \infty$.  Consequently, exactly one scale factor
{\em vanishes} while the remaining two {\em diverge} as $|\tau| \to
\infty$.  Thus if one begins with a cubical cell at some finite $\tau$,
then the cell will become {\em planar} after both forward and backward
evolution (under Kasner evolution the cell will become planar in
`future' and one dimensional in the `past)'. Since $\tau \to \pm \infty$
correspond to $t \to \pm \infty$, the vanishing/diverging behaviour of
scale factors never occurs for {\em finite t}.  The modified solution is
thus {\em non-singular}.

Consider the behaviour of the volume, $V^2 = p^1 p^2 p^3$. For $|\tau|
\to \infty, V^2 \to \Case{\epsilon^3}{64} |\alpha_1 \alpha_2 \alpha_3|
e^{2 |\tau|}$, and therefore it must have a minimum, $V_*$, for some
$\tau_*$.  It is clear that for $\tau$ larger (smaller) than all
$(\tau_I)_0$, $V^2$ is monotonic. Thus $\tau_*$ must lie in the interval
of the minimum and maximum values of the parameters $(\tau_I)_0$. The
minimum volume, $V_*$, reached by any {\em particular} solution depends
on $\vec{\alpha}$ as well as on $\vec{\tau}_0$.  For any given
$\vec{\alpha}$, the smallest possible minimum volume is attained for
solutions for which all $(\tau_I)_0$ are equal.  Parameterizing the
$\vec{\alpha}$ as \cite{BKL}: $\alpha_1 := -u /(1 + u + u^2), \alpha_2
:= (1 + u)/(1 + u + u^2), \alpha_3 := (u + u^2)/(1 + u + u^2), u \ge 1$,
this smallest possible minimum volume for a given $u$, is given by, 
\begin{equation} 
(V_*)_{\text{min}} =
\sqrt{\Case{\epsilon^3 |\alpha_1 \alpha_2 \alpha_3|}{8}}  = 
\left(\frac{\epsilon}{2}\right)^{\frac{3}{2}} \left[\frac{u (1 + u)}{\left(1 + u +
u^2\right)^{\frac{3}{2}}}\right] 
\end{equation} 
which can be arbitrarily small though strictly positive. 

Since singularity theorem is evaded, `energy' condition(s) must be
violated. In the present context, this means that $R_{tt} < 0$ must hold
{\em some where} in the space-time. By computing $R_{tt} = - \sum_I
a_I^{-1}\frac{d^2 a_I}{dt^2}$, one can see easily that this is so in the
neighbourhood of $V_*$. 

So far we have focussed on the features of the exact solution.  Recall
that for the validity of the effective Hamiltonian we also need $p^I
\gtrsim q$, or in the dimensionless variables used above, $p^I \gtrsim
1$. This puts a {\em restriction on the time $\tau$ for which any
specific solution can be trusted as an approximation.} Specifically,
cosh$\left\{ (1 - \alpha_I)(\tau - (\tau_I)_0) \right\} \gtrsim
\Case{2}{\epsilon |\alpha_I|}$. Had we used the effective Hamiltonian
obtained from the continuum approximation \cite{Spin,BIX} which is same
as the Einstein Hamiltonian, we would get the usual Kasner solution and
obtain the restriction on $\tau$ as, exp $\left\{ (1 - \alpha_I)(\tau -
(\tau_I)_0) \right\} \gtrsim \Case{4}{\epsilon |\alpha_I|}$. For ease of
comparison, we have just written $p_0^I := \Case{\epsilon |\alpha_I|}{4}
\text{exp} \left\{ - (1 - \alpha_I)(\tau_I)_0 \right\}$ so that at for
{\em large} $\tau$, both the modified and the usual Kasner solutions
match.  Since the modified solution is always larger than the Kasner
solution at the same $\tau$, the $\tau-$regime of validity for the
effective solution is {\em larger} than that for the Kasner solution.
However, since $\Case{2}{\epsilon |\alpha_I|} > \Case{2}{\epsilon} > 2$,
the parameters $(\tau_I)_0$ are out side the domain of validity and so
is the smallest (non-zero) value of triad components. The `bounce' in
the triad components is thus in the {\em quantum regime}. This is {\em
different} from the isotropic case \cite{Bounce} where the bounce
implied by effective dynamics {\em can} occur in the domain of validity
of the effective Hamiltonian, depending upon details of matter
Hamiltonian. Thus, in the simplest of the anisotropic models, although the
effective dynamics inferred {\em is} non-singular due to bouncing triad
components, these bounces lie in the quantum domain and for a `reliable'
removal of singularity, one still needs to appeal to the quantum theory.

In summary, we make three points: (1) the method of effective
Hamiltonian can be extended to homogeneous, anisotropic models and leads
to {\em non-singular} effective dynamics with the exact solution
parameterized by the same Kasner parameters; (2) unlike (at least some
of) the isotropic models, for reliable singularity removal, one has to
appeal to quantum theory; and (3) the quantum theory stipulates
modifications of the  classical Einstein dynamics, not only for small
triads ($p^I$) but also for larger values of their conjugates ($K_I$)
which is responsible for the non-singularity. 

\begin{acknowledgments}
I would like to thank Martin Bojowald and Golam Hossain for useful
remarks. This work was initiated during my visit to AEI, Golm during
December 2004. The warm hospitality is gratefully acknowledged.
\end{acknowledgments}

\end{document}